\documentclass[fleqn,preprint,showpacs,preprintnumbers,amsmath,amssymb]{revtex4}
\usepackage{graphicx}
\usepackage{dcolumn}
\usepackage{bm}

\begin{document}

\title{Quantum collapse in ground-state Fermi-Dirac-Landau plasmas}
\author{M. Akbari-Moghanjoughi}
\affiliation{Azarbaijan University of
Tarbiat Moallem, Faculty of Sciences,
Department of Physics, 51745-406, Tabriz, Iran}

\date{\today}
\begin{abstract}
It is revealed that in a relativistically degenerate dense highly-magnetized electron-ion plasma the effective quantum-potential due to the total quantum-force acting on fermions may cancel-out causing a quantum transverse collapse in the ground-state Fermi-Dirac-Landau (GSFDL) plasma. The condition for the plasma transverse collapse is found to be restricted to the minimum relativistic degeneracy parameter and minimum impressed magnetic field strength values satisfied for many superdense astrophysical objects such as white dwarfs and neutron stars. In such plasmas, the magnetization pressure is shown to cancel the lateral electron degeneracy pressure counteracting the existing gravitational pressure. Furthermore, using the Sagdeev pseudopotential method in the framework of quantum magnetohydrodynamics (QMHD) model including spin magnetization it is confirmed that the quantum pressure due to spin-orbit polarization and the electron relativistic degeneracy has significant effects on the existence criteria and the propagation of localized magnetosonic density excitations in GSFDL plasmas. Current findings can have important implications for the density excitations mechanism and gravitational collapse of the highly magnetized astrophysical relativistically dense objects such as white-dwarfs, neutron stars, magnetars and pulsars.
\end{abstract}

\keywords{Fermi-Dirac-Landau plasmas, Relativistic degeneracy, Relativistic Paramagnetism and Diamagnetism, Spin-orbit Magnetization, Magnetoacoustic nonlinear wave, Quantum magnetohydrodynamics, Sagdeev potential}
\pacs{52.30.Ex, 52.35.-g, 52.35.Fp, 52.35.Mw}
\maketitle

\section{Introduction}

There has been a great interest towards the study of thermodynamics and nonlinear properties of a degenerate dense magnetized matter, since the early works of Chandrasekhar, Bohm, Pines and Levine et.al \cite{chandra1, bohm, pines, levine}. In a degenerate cold and dense plasma, the so-called quantum plasma state, interesting quantum effects arise due to the fundamental Pauli exclusion rule on the dynamics of degenerated species \cite{landau}. Quantum tunneling, electron degeneracy pressure, quantum paramagnetism etc. are some of the well-known properties of metallic and semiconductor nano-structured materials such as, nano-particles, quantum-wells, quantum-wires and quantum-dots \cite{haug} which are of great interest in the fast growing fields of quantum optics, electronics and computation. There are also a great majority of recent activities in the field of quantum hydrodynamics (QHD) and quantum magnetohydrodynamics (QMHD) \cite{haas1, haas2, akbari1, akbari2, Markowich, Marklund1, Brodin1, Marklund2, Brodin2, manfredi, shukla, gardner} which highlight the unexpected nonlinear wave features in quantum plasmas. Particularly, it has been shown that the thermodynamical properties of a degenerate gas may be different when the fermions become relativistic \cite{kothary}. Recently, it has been confirmed that the nonlinear wave properties of a degenerate plasma also substantially differ in relativistically degenerate plasma from those of an ordinary quantum plasma \cite{akbari3, akbari4}.

Interesting phenomenon occur when the magnetic field is involved. Marlkund and Brodin have recently extended the QHD model to include the spin-magnetization effects \cite{Marklund3} by introducing a generalized term for the so-called quantum force. Their initial studies indicate that in a perfectly conductive quantum plasma, the spin magnetization can act as a negative pressure-like entity significantly modifying the dynamics of the spin-induced nonlinear magnetosonic waves \cite{Marklund4}. Many recent studies using the spin-included QMHD model also justify the dominant spin effects on nonlinear wave properties in quantum plasmas \cite{misra1, martin, mushtaq, brodin3, zaman, vitaly, misra2, akbari5}. Because of the existence of very high magnetic fields and relativistic degeneracy effects in astrophysical dense objects such as white dwarfs and neutron stars, etc. and the related origin of star collapse to these effects, the subject of a dense matter in high magnetic field has been studied extensively for many decades \cite{weeler}. Chandrasekhar \cite{chandra2, chandra3} has shown that the relativistic degeneracy of Fermi-Dirac electrons can give rise to a collapse of a dense degenerate star. However, the coexistence of large magnetic fields in collapsing stars did not seem a coincidence.

In a series of papers Canuto and Chiu \cite{can1, can2, can3, can4, can5, can6, can7, can8} have thoroughly investigated the behavior of Fermi-Dirac relativistically degenerate gas in arbitrarily high magnetic field and found very interesting results. They showed that, the thermodynamics and magnetic properties of a relativistically degenerate Fermi gas may differ substantially regarding to those of a normally degenerate ones. In particular, they proved that unlike an ordinary degenerate matter when the electrons become relativistic the magnetic susceptibility can show oscillatory behavior for very high electron number-densities. This is because of the landau orbital contribution to the magnetization of the gas which becomes dominant for higher electron number-densities. These findings triggered a notion that the large magnetic field present in some compact astrophysical degenerate objects may in fact play a key role in the collapse phenomenon observed for some white-dwarfs and neutron-stars. O'Connell and Roussel \cite{con1} have shown that a Landau orbital ferromagnetism (the so-called LOFER) state solution exist in both nonrelativistic and relativistically degenerate magnetized Fermi-gas which can have direct consequences on the origin of a large magnetic fields present in astrophysical compact objects such as white-dwarfs and neutron stars\cite{con2}.

On the other hand, more recently, Chaichian et al. \cite{chai} have investigated the possibility of a transverse magnetic collapse in a relativistically degenerate Fermi-Dirac gas showing that the gas can experience an anisotropic pressure normal to the magnetic field and the magnetization pressure may even cancel the degeneracy pressure leading to a collapse in equatorial plane of plasma. \textbf{However, as pointed-out by Blandford and Hernquist \cite{bland}, the pressure is isotropic, contrary to what is stated by Canuto and Ventura \cite{can9} and when compressing the electron gas perpendicular to the field, ${\bf{B}}$, a work must be done against the Lorentz force density, ${\bf{J_{m}}} \times {\bf{B}}$ (${\bf{J_{m}}}$, being the magnetization-current density), involving the magnetization current. Thus, there is a magnetic contribution to the perpendicular pressure which may be the key to the origin of giant magnetic field present in astrophysical compact objects like white-dwarfs and neutron stars. Dong Lai \cite{dong} has given a complete review on different aspects of matter in strong magnetic fields.}

\textbf{In current investigation, using a complete quantum magnetohydrodynamics model which includes the effect of magnetization current, we attempt to explore the effect of high degeneracy pressure and magnetic field on wave dynamics of Ground-State Fermi-Dirac-Landau (GSFDL) plasmas, using the spin-included QMHD model to show that a quantum collapse in such plasma is still possible in the presence of the Lorentz magnetic force. The presentation of the paper is as follows. The QMHD plasma model including the spin contribution is introduced in Sec. \ref{equations} and the Localized magnetosonic density solution is given in Sec. \ref{calculation}. The numerical results are presented in Sec. \ref{discussion} and the concluding remarks are given on Sec. \ref{conclusion}.}

\section{QMHD Model including Spin-Magnetization Effects}\label{equations}

\textbf{The basic QMHD equation set including the magnetization, Lorentz-force and quantum tunneling effects in a center of mass frame include the continuity equation
\begin{equation}\label{cont}
\frac{{\partial \rho_c }}{{\partial t}} + \nabla  \cdot (\rho_c {\bf{U_c}}) = 0,
\end{equation}
where, $\rho_c=m_e N_e + m_i N_i$ is the center of mass plasma density and $\bm{U_c}=(N_e m_e \bm{U_e} + N_i m_i \bm{U_i})/\rho_c$ is the center of mass velocity. Also, the extended momentum equation \cite{Marklund4}
\begin{equation}\label{mom}
\frac{{\partial {\bf{U_c}}}}{{\partial t}} + \left( {{\bf{U_c}} \cdot \nabla } \right){\bf{U_c}} = {\rho_c ^{ - 1}}\left( {{\bf{J_{m}}} \times {\bf{B}} - \nabla P_c + {{\bf{F}}_Q}} \right),
\end{equation}
where, the quantum force $\bm{F}_Q={\bm{F}}_B+{\bm{F}}_m$ is a collective contribution of quantum Bohm-force, ${\bm{F}}_B$, and the magnetization-force, ${\bm{F}}_m$, namely,
\begin{equation}\label{QF}
{{\bf{F}}_Q} = \frac{{{\rho_c\hbar ^2}}}{{2{m_e}{m_i}}}\nabla \frac{{\Delta \sqrt {{\rho _c}} }}{{\sqrt {{\rho _c}} }} + {{\Gamma}} \nabla {{B}}.
\end{equation}
The quantities $\bf{J_{m}}$, $\bf{B}$ and $P_c$ are the magnetization current, the magnetic field with, $\Gamma$, being the plasma magnetization per unit volume, and scalar center of mass pressure, respectively. The magnetization current, $\bf{J_{m}}$, is related to the plasma magnetization vector through the relation; ${\bf{J_{m}}} = \mu _0^{ - 1}\nabla  \times ({\bf{B}} - {\mu _0}{\bf{\Gamma}})$.}

Now, we consider an ideal collisionless, perfectly conducting and completely degenerate electron-ion plasma, subject to a uniform arbitrarily-high magnetic field, $\textbf{B}=B_0 \bf{\hat k}$. By ignoring the local variations of magnetic field in plasma compared to the very strong external-field, the simplified QMHD equation set including the Bohm force, the electron degeneracy pressure and spin-orbit magnetization effect in the center of mass frame, read as
\begin{equation}\label{basic}
\begin{array}{l}
\frac{{\partial N_i }}{{\partial t}} + \nabla  \cdot (N_i {\bf{U_c}}) = 0,\\
\frac{{\partial {{\bf{U}}_{\bf{c}}}}}{{\partial t}} + \left( {{{\bf{U}}_{\bf{c}}}\cdot\nabla } \right){{\bf{U}}_{\bf{c}}} = {\rho_c}^{-1}\left[ {\nabla (\Gamma B_0) - \nabla {P} + \frac{{{\rho_c}{\hbar ^2}}}{{2{m_e m_i}}}\nabla \frac{{\Delta \sqrt {{N_i}} }}{{\sqrt {{N_i}} }}} \right],
\end{array}
\end{equation}
\textbf{where, we have employed $\rho_c\simeq N_i m_i$ from $N_e\simeq N_i=N$, i.e., quasineutrality condition. Note also that, the total plasma pressure is the combination of the zero-field electron degeneracy-pressure plus the ground-state transverse-pressure element due to magnetization, i.e., $P=P_{B=0}+P(l,B,N)$. As will be shown later, the $P_{B=0}$ part of the pressure is an inevitable contribution without which no propagations will be possible in GSFDL plasma. In our particular case, we are interested in the state of magnetization where electrons are populated in the ground Landau level or strongly quantized GSFDL plasma \cite{dong} (i.e., $l$=$0$, $l$ being the Landau level). Therefore, the ground-state plasma magnetization which is due to both Pauli electron spin- and Landau electron orbital-polarizations in a magnetized relativistically degeneracy Fermi-gas, which according to Canuto and Chiu \cite{can2} can not be separated into paramagnetic and diamagnetic parts, may be written in the following form
\begin{equation}\label{QP}
{\Gamma}(\eta) = \frac{{{\mu _B}}}{{2{\pi ^2}\mathchar'26\mkern-10mu\lambda_c^{3}}}\left\{ {\eta \sqrt {1 + {\eta ^2}}  - \ln \left[ {\eta  + \eta \sqrt {1 + {\eta ^2}} } \right]} \right\},
\end{equation}
in which $\mu_B=e\hbar/m_ec$ is the Bohr magneton and $\mathchar'26\mkern-10mu\lambda_c=\hbar/m_ec$ is the Compton wavelength. The normalized parameter $\eta=(3/2)(N/N_c)(B_c/B_0)$ is defined as the fraction of electron number-density to magnetic field strength each with the normalization value, $B_c=m_e^{3}c^2/e\hbar\simeq 4.41\times 10^{13}G$ and ${N_c} = {{8\pi m_e^{3}{c^3}}}/{{3{h^3}}}\simeq 5.9 \times 10^{29} cm^{-3}$, respectively. It has been shown that \cite{can2} the susceptibility of a magnetized Fermi gas reduces to the Curie-Langevin law in the weak-quantization (low density and high temperature \cite{dong}) limit. It has also been found that, by increasing the electron number-densities and by filling the higher Landau-levels the magnetization passes through many maximums and minimums \cite{can2}. As it is shown in Fig. 1, as the field strengths is lowered the higher Landau levels can become available to electrons leading to the "de Haas-van Alphen effect like" oscillatory behavior of the magnetization.}

\textbf{Moreover, the thermodynamic pressure of a relativistically degenerate Fermi-gas in the absence of a magnetic-field, i.e. $P_{B=0}$, is given by \cite{chandra1}
\begin{equation}\label{p}
{P_d} = \frac{{\pi m_e^4{c^5}}}{{3{h^3}}}\left[ {R\left( {2{R^2} - 3} \right)\sqrt {1 + {R^2}}  + 3{{\sinh}^{ - 1}}R} \right],
\end{equation}
in which, the relativity parameter, $R=P_{Fe}/(m_e c)=(N/N_c)^{1/3}$ is a measure of the relativistic degeneracy and $P_{Fe}$ is the electron Fermi relativistic momentum. Note that, in the limits of very small relativity parameter, $R\ll R_{Ch}$ (the value $R_{Ch}=\sqrt{2}$ being the Chandrasekhar critical relativistic degeneracy value), a degeneracy pressure of the form $P_d={{{\left( {3/\pi } \right)}^{2/3}}{h^2}N_e^{5/3}/(20{m_e})}$ and for corresponding very large values, $R\gg R_{Ch}$, a form $P_{d}={{{\left( {3/\pi } \right)}^{1/3}}hcN_e^{4/3}}/8$ is expected for the electron degeneracy pressure. It is observed from Fig. 1 that, for a fixed magnetic field strength, as the relativity parameter and consequently the electron number-density increases the electrons occupy the higher Landau levels and the magnetization passes through several maximum an minimum values. It can be shown that the maximum values of the magnetization, as separated by thin vertical lines, take place at $R\simeq ({2l})^{1/6}\sqrt{B}$ or $B\simeq (2l)^{-2/6}R^2$, respectively, for Figs. 1(a) and 1(b).}

For the propagation of magnetoacoustic nonlinear waves along the $x$ direction perpendicular to the constant magnetic field $B_0$ along the $z$-axis, we derive
\begin{equation}\label{dimensional}
\begin{array}{l}
\frac{{\partial N}}{{\partial t}} + \frac{{\partial N{U_c}}}{{\partial x}} = 0, \\
\frac{{\partial {U_c}}}{{\partial t}} + {U_c}\frac{{\partial {U_c}}}{{\partial x}} =  - \frac{{{m_e}{c^2}}}{{{m_i}}}\frac{\partial }{{\partial x}}\left( {\int {\frac{{{d_R}{P_d}(R)}}{R}} dR} \right) + \frac{{3{m_e}{c^2}}}{{4{m_i}}}\frac{\partial }{{\partial x}}\left( {\int {\frac{{{d_\eta }\Gamma (\eta )}}{\eta }} d\eta } \right)
+ \frac{{{\hbar ^2}}}{{2{m_e}{m_i}}}\frac{\partial }{{\partial x}}\left( {\frac{1}{{\sqrt N }}\frac{{{\partial ^2}\sqrt N }}{{\partial {x^2}}}} \right). \\
\end{array}
\end{equation}
It is convenient, thus, to define an effective quantum potential in the form
\begin{equation}\label{dimensional}
{\Psi _{eq}} = \frac{3}{4}\int {\frac{{{d_\eta }{\Gamma}(\eta )}}{\eta }} d\eta  - \int {\frac{{{d_R}{P_d}(R)}}{R}} dR, \\
\end{equation}
where, $\eta=3R_0^{3}/2\bar B$ ($\bar B=B_0/B_c$) and $R=R_0 (N/N_0)^{1/3}=R_0 \bar{n}^{1/3}$ with $R_0=(N_0/N_c)^{1/3}$. In order to obtain a dimensionless set of equations we may use the following scalings
\begin{equation}\label{normal}
x \to \frac{{{c_{s}}}}{{{\omega _{pi}}}}\bar x,\hspace{3mm}t \to \frac{{\bar t}}{{{\omega _{pi}}}},\hspace{3mm}N \to \bar n{N_0},\hspace{3mm}U_c \to \bar u{C_{s}}.
\end{equation}
In the normalizing factors the parameters, $N_0$, ${\omega _{pi}} = \sqrt {{e^2}{N_{0}}/(\varepsilon_0{m_i})}$ and ${C_{s}} = c\sqrt {{m_e}/{m_i}}$ denote the equilibrium electron number-density, characteristic ion plasma frequency and ion sound-speed (despite the name this speed may be comparable to the Fermi-speed of an electron in a typical metal), respectively. Hence, we obtain
\begin{equation}\label{dimensionless}
\begin{array}{l}
\frac{{\partial \bar n}}{{\partial t}} + \frac{{\partial \bar n{\bar u}}}{{\partial x}} = 0, \\
\frac{{\partial {\bar u}}}{{\partial t}} + {\bar u}\frac{{\partial {\bar u}}}{{\partial x}} = \frac{{\partial {\Psi _{eq}}}}{{\partial x}} + {H^2}\frac{\partial }{{\partial x}}\left( {\frac{1}{{\sqrt {\bar n} }}\frac{{{\partial ^2}\sqrt {\bar n} }}{{\partial {x^2}}}} \right), \\
{\Psi _{eq}} = \frac{3}{2}\sqrt {1 + {\eta ^2}}  - \sqrt {1 + {R^2}}. \\
\end{array}
\end{equation}
where, we have defined a new parameter called the quantum diffraction parameter as, $H = \sqrt {{m_i}/{2m_e}} (\hbar {\omega _{pi}})/({m_e}{c^2}) = e\hbar \sqrt {{N_c}R_0^3/\pi } /(2m_e^{3/2}{c^2})$. In the forthcoming algebra, the bar notations here and after are avoided for clarity. The gradient of the effective quantum potential \cite{akbari6} is the local quantum force which is central to our analysis of the quantum-collapse and is the sum of the quantum-forces due to relativistic electron degeneracy pressure and the relativistic spin-orbit magnetization. It is really interesting that the two potentials are of the same shape and order of magnitude, hence, there may exist points at which the total effective quantum-potential vanishes.

\section{Localized Magnetosonic Excitations}\label{calculation}

\textbf{In order to study the localized excitations in the co-moving frame, we first employ the coordinate transformation of the form, $\xi=x-M t$, to Eqs. (\ref{dimensionless}) where, $M=U_c/C_s$ is the Mach-number of the nonlinear excitations. Furthermore, the integration of the continuity equation with respect to $\xi$ with boundary conditions $\mathop {\lim }\limits_{\xi  \to  \pm \infty } \bar n = 1$ and $\mathop {\lim }\limits_{\xi  \to  \pm \infty } \bar u = 0$ leads to $u = M\left( {{1}/{n} - 1} \right)$. By replacing this value for, $u$, in the momentum equation, using the change of variable, $\bar n=Z^2$, and integrating with the same boundary conditions, the Eqs. (\ref{dimensionless}) can be reduced to the following nonlinear differential equation
\begin{equation}\label{diff2}
\begin{array}{l}
\frac{{{H^2}}}{Z}\frac{{{\partial ^2}Z}}{{\partial \xi }} = \frac{{{M^2}}}{2}{\left( {1 - {Z^{ - 2}}} \right)^2} - {M^2}\left( {1 - {Z^{ - 2}}} \right) + \sqrt {1 + R_0^2{Z^{4/3}}}  \\ - \sqrt {1 + R_0^2}  - (3/2)\sqrt {1 + 9R_0^6{Z^{4}}/({4B^2})}  + (3/2)\sqrt {1 + 9R_0^6/({4B^2})}. \\
\end{array}
\end{equation}
Multiplying both sides of Eq. (\ref{diff2}) by $dZ/d\xi$ and integrating with appropriate boundary conditions (mentioned above), we arrive at the well known energy integral of the form in terms of the number-density variable}
\begin{equation}\label{energy}
{({d_\xi } n)^2}/2 + U(n) = 0,
\end{equation}
with the pseudopotential given as
\begin{equation}\label{pseudo}
\begin{array}{l}
U(n) = \frac{n}{{4{H^2}}}\left[ {\frac{{4{M^2}}}{n} + 4{M^2}n - 2\sqrt {1 + R_0^2}  + 8n\sqrt {1 + R_0^2}  - 8{M^2} + \frac{{3\sqrt {1 + R_0^2} }}{{R_0^2}}} \right. \\
- 6n\sqrt {1 + {n^{2/3}}R_0^2}  - \frac{{3{n^{1/3}}\sqrt {1 + {n^{2/3}}R_0^2} }}{{R_0^2}} + 6\sqrt {1 + \frac{{9R_0^6}}{{4{B^2}}}}  - 12n\sqrt {1 + \frac{{9R_0^6}}{{4{B^2}}}}  + 6n\sqrt {1 + \frac{{9{n^2}R_0^6}}{{4{B^2}}}}  \\ \left. { - \frac{3}{{R_0^3}}{{\sinh }^{ - 1}}{R_0} + \frac{3}{{R_0^3}}{{\sinh }^{ - 1}}\left( {{n^{1/3}}{R_0}} \right) - \frac{{4B}}{{R_0^3}}{{\sinh }^{ - 1}}\left( {\frac{{3R_0^3}}{{2B}}} \right) + \frac{{4B}}{{R_0^3}}{{\sinh }^{ - 1}}\left( {\frac{{3nR_0^3}}{{2B}}} \right)} \right]. \\
\end{array}
\end{equation}
We now evaluate the existence of the nonlinear localized density structures which requires that the following conditions to satisfy, simultaneously
\begin{equation}\label{conditions}
{\left. {U(n)} \right|_{n = 1}} = {\left. {\frac{{dU(n)}}{{dn}}} \right|_{n = 1}} = 0,\hspace{3mm}{\left. {\frac{{{d^2}U(n)}}{{d{n^2}}}} \right|_{n = 1}} < 0.
\end{equation}
Another condition must be met in order for the exclusion of shock-like structures such that, for at least one either maximum or minimum nonzero $n$-value, we have $U(n_{m})=0$, so that for every value of $n$ in the range ${n _m} > n  > 1$ (compressive soliton) or ${n _m} < n  < 1$ (rarefactive soliton), $U(n)$ is negative (it is understood that there is no root in the range $[1,n_m]$). In such a condition we can obtain a potential minimum which describes the possibility of a solitary wave propagation. It should be noted that the both maximum and minimum $n$-value may coexist which is a case for the double-well, namely, the coexistence of the compressive and rarefactive solitary structures \cite{akbari7}.

The general solution of the pseudo-potential satisfying the above conditions may be written as
\begin{equation}\label{soliton}
\xi  - {\xi _0} =  \pm \int_1^{n_m} {\frac{{dn}}{{\sqrt { - 2U(n)} }}}.
\end{equation}
Obviously the first two conditions in Eq. (\ref{conditions}) is met for the pseudo-potential given by Eq. (\ref{pseudo}). Concerning the third condition, one obtains from direct evaluation of the second derivative of the Sagdeev potential, Eq. (\ref{pseudo}), at the unstable point, $n=1$,
\begin{equation}\label{dd}
{\left. {\frac{{{d^2}U(n)}}{{d{n^2}}}} \right|_{n = 1}} = \frac{2}{{{H^2}}}\left[ {{M^2} - \frac{{R_0^2}}{{3\sqrt {1 + R_0^2} }} + \frac{{81R_0^6}}{{12{B^2}\sqrt {1 + 9R_0^6/(4{B^2})} }}} \right].
\end{equation}
This requires that, for existence of a magnetoacoustic solitary excitation in Fermi-Dirac-Landau plasma, the soliton matching speed should be below a critical value defined as
\begin{equation}\label{consol2}
{M_{cr}} = \frac{1}{2}\sqrt {\frac{{4R_0^2}}{{3\sqrt {1 + R_0^2} }} - \frac{{27R_0^6}}{{B\sqrt {4{B^2} + 9R_0^6} }}}.
\end{equation}
\textbf{It is interesting to note that, when the first term in the square root which corresponds to the electron degeneracy pressure is absent, the solitary excitation will not be possible in the plasma. The critical Mach-number value obtained here, disregarding the magnetic part, agrees well with the value estimated for an unmagnetized Fermi-Dirac plasma \cite{akbari8}. Now, let us examine the existence of the $n_m$ values which is essential to our analysis. As it was mentioned above we require that for at least one either maximum or minimum nonzero $n_m$-value, we must have $U(n_m)=0$, so that for every value of $n$ in the range ${n_m} > n  > 0$ or ${n_m} < n  < 0$, $U(n)$ is negative. In such a condition there will be a potential minimum which describes the propagation of solitary nonlinear structure.} To do this we inspect the limits of the pseudopotential given in Eq. (\ref{pseudo}) which are given below
\begin{equation}\label{nm}
\mathop {\lim }\limits_{n \to 0} U(n) = \frac{{{M^2}}}{{{H^2}}} > 0,\hspace{3mm}\mathop {\lim }\limits_{n \to \infty } U(n) =  + \infty.
\end{equation}
\textbf{As it is clearly evident the above result is independent from GSFDL plasma parameters. Therefore, it is confirmed that, a solitary double-well and periodic double-layer are possible in the plasma under consideration. The possibility of such structures has already been confirmed in an unmagnetized Fermi-Dirac plasma \cite{akbari6}. A closer inspection of Eqs. (\ref{consol2}) and (\ref{nm}) reveals that in the absence of degeneracy pressure even periodic nonlinear excitations will vanish. This may be a justification for the addition of a zero-field part in the perpendicular pressure component defined earlier.}

\section{Numerical Analysis and Discussion}\label{discussion}

\textbf{As has been mentioned in the previous section, the effective quantum potential, $\Psi_{eq}$ defined in Eq. (\ref{dimensionless}), may vanish for different choices of the values of relativistic degeneracy parameter, $R$, and the normalized magnetic field strength, $B$, i.e., the lateral electron degeneracy-pressure can be totaly canceled by the corresponding magnetization pressure at some given values of magnetic field. This effect can lead to a phenomenon called the quantum collapse of plasma in plane perpendicular to the magnetic field if a gravity force acts from outside. Figure 2(a) shows a curve in $R$-$B$ plane where the effective quantum potential vanishes (or correspondingly where the plasma undergoes a transverse quantum collapse). However, by solving the equation $\Psi_{eq}=0$, it can be easily shown that the effective quantum potential does not vanish bellow the minimum values of $B_m=135\sqrt{3}/32$ and $R_m=\sqrt{5}/2$. It is also noted that, the magnetic field strength required ($B\ge B_m$) to completely cancel the lateral quantum-pressure is extremely high (i.e. $B_m\simeq32.2\times10^9T$). However, this high value of magnetic field may be expected to exist in some superdense astrophysical compact objects. It is interesting that the lowest value of required field ($B_m$) for quantum collapse corresponds to the relativistic degeneracy value of $R=\sqrt{15/8}$ which is very close to the Chandrasekhar value of $R_{Ch}=\sqrt{2}$ for the gravitational collapse. Obviously there are many other effects such as the plasma rotation \cite{akbari6}, electron-electron interactions and finite temperature effects which can modify the value of $B_m$. Furthermore, for all the values of $B>B_m$ there will be two values of relativistic degeneracy at which the effective quantum potential vanishes. In Fig. 1(b) the variation of $\Psi_{eq}$ is shown with respect to the change in the relativity parameter for various applied magnetic field strengths. It is revealed that always a minimum value for the effective potential exists which is due to the delicate interplay between the magnetization and the electron degeneracy effects. This can be an indication of the GSFDL plasma stability after the first quantum collapse. However, for stronger magnetic fields ($B>B_m$), as the relativity parameter is increased, the GSFDL plasma may undergo another collapse, i.e., the potential vanishes at some other relativity parameter values the value of which increases as the strength of the magnetic field is increased. Solving the equation $\Psi_{eq}=0$, after some algebra, we can obtain an analytical expression for the values of $R$ at which the effective potential vanishes for a given value of the magnetic field strength, $B$. These values are given as
\begin{equation}\label{R1}
{R_{1,2}} = {\left[ {\frac{{8{{( \mp 6)}^{2/3}}{B^2} \mp {{( \mp 6)}^{1/3}}{{\left( {\sqrt {164025{B^4} - 3072{B^6}}  - 405{B^2}} \right)}^{2/3}}}}{{27{{\left( {\sqrt {164025{B^4} - 3072{B^6}}  - 405{B^2}} \right)}^{1/3}}}}} \right]^{\frac{1}{2}}}.
\end{equation}
As an example for the ideal GSFDL plasma such as a neutron star in a magnetic field of order $10^{15}G$, we get $R_{1}\simeq1.25$ and $R_{2}\simeq43.8$ or equivalently the quantum-collapse occurs at electron number-densities of $n_{01}\simeq1.15\times 10^{30}cm^{-3}$ and $n_{02}\simeq4.96\times 10^{34}cm^{-3}$. When the effective potential vanishes the plasma becomes thermodynamically unstable and complex and the existence of the localized density excitation can not be studied using the model employed here. However, it can be shown that at the minimum of the effective potential the critical soliton Mach-number vanishes, hence, the solitary excitations can occur only for the relativity parameter values below the corresponding minimum value of the effective potential value, that is, there is cut-off value of relativity for solitary excitations above which the localized density excitations will not be present.}

\textbf{Figure 3 depicts the volume of occurrence for localized density excitations in $M$-$B$-$R_0$ space. It is confirmed that, in accordance with the above statement, for every value of field strength, $B$, the possible Mach-number values are limited from above by a cut-off value of relativity parameter value, which corresponds to the minimum of the effective quantum potential shown in Fig. 1(b). The soliton Mach-numbers for fixed magnetic field strength is remarked to have always a maximum, value of which, increases with increase in the magnitude of the applied magnetic field.}

\textbf{Moreover, Fig. 4 depicts the variation of pseudopotential profiles (Eq. \ref{pseudo}) with respect to different fractional plasma parameter when other parameters are fixed. It is evident that in the GSFDL plasma under consideration double-wells and double-layers can exist. This effect should not be confused with the similar pseudopotential shapes presented in Ref. \cite{akbari7}, which are due to the existence of dust ingredients. It is further revealed that increase in the value of the relativity parameter (plasma relativistic degeneracy value) increases/decreases the amplitude of the raefactive/compressive solitons (Fig. 4(a)), while, increase in strength of the magnetic field has the converse effect (Fig. 4(b)). The rarefactive periodic double-layers are shown in Figs. 4(a) and 4(b). The effect of the variation in Mach-number (Fig. 3(c)) on soliton amplitude indicates that the soliton with higher amplitude propagate with the slower Mach values. This effect is identical on both rarefactive and compressive solitary structures shown as double-wells. Finally, Fig. 4(d) reveals that large amplitude compressive periodic double-layer are also possible in GSFDL plasmas.}

\textbf{Figure 5 shows different regions for Landau-levels (the regions, where, $\sqrt {1 + 2lB}  < \sqrt {1 + 9{R^6}/4{B^2}}  < \sqrt {1 + 2(l + 1)B}$ with $l$ being the Landau-level) \cite{can2}. It is confirmed that quantum collapse, $\Psi_{eq} =0$, resides totaly in the ground-state ($l$=$0$) region, hence the name, GSFDL plasma.}

\section{Summary}\label{conclusion}

\textbf{Quantum magnetohydrodynamics model has been used to show that in a ground-state Fermi-Dirac-Landau plasma the local quantum force due to electron relativistic degeneracy and the spin-orbit polarization magnetization have the same order of magnitude but different signs, hence, the effective quantum local force may vanish leading to quantum collapse of plasma perpendicular to the applied magnetic field. It was revealed that, the condition for quantum collapse is satisfied for some astrophysical dense objects and is limited to the ground-state plasma. It was further shown that, for a given applied magnetic field strength, a cut-off value of relativity parameter (or equivalently the electron number-density) exists above which no localized magnetosonic structure can form. Current findings can have important implications for the density excitations and collapse of the highly magnetized of astrophysical relativistically dense objects such as white-dwarfs, neutron stars, magnetars and pulsars.
}

\newpage

\textbf{FIGURE CAPTIONS}

\bigskip

Figure-1

\bigskip

(Color online) Variation of the plasma magnetization (thick blue-line) with respect to increase in relativity parameter and magnetic field strength. The vertical axis ($\Gamma$) indicates the magnetization due to both electron spin and orbit contributions. The Landau levels are separated by thin vertical lines and for Fig. 1 (left plot) fill from left starting $l$=$0$ (light-blue color) and for Fig. 1 (right plot) fill from right.

\bigskip

Figure-2

\bigskip

(a) The possibility of zero effective potential in GSFDL plasma. (b) The variation of effective quantum potential with respect to the relativistic degeneracy parameter, $R$ and normalized magnetic impressed-field strength, $B$.

\bigskip

Figure-3

\bigskip

Figure 3 shows a volume in $M$-$B$-$R_0$ space in which a solitary excitation can exist.

\bigskip

Figure-4

\bigskip

Variation of the Sagdeev pseudopotential profile with respect to change of different plasma fractional parameter values when the others are fixed. The size of the dashes in each plot changes according to the change in the value of the varied parameter.

\bigskip

Figure-5

\bigskip

(Color online) Figure 5 shows different Landau-level regions ($\sqrt {1 + 2lB}  < \sqrt {1 + 9{R^6}/4{B^2}}  < \sqrt {1 + 2(l + 1)B}$ with $l$ being the Landau state) in the $B$-$R$ plane confirming that the magnetic collapse completely resides in the ground-state region (largest colored region) of a Fermi-Dirac-Landau plasma and is bounded from left and below by the minimum values of $R_m=\sqrt{5}/2$ and $B_m=135\sqrt{3}/32$.

\bigskip

\newpage

\begin{figure}[ptb]\label{Figure1}
\includegraphics[scale=.6]{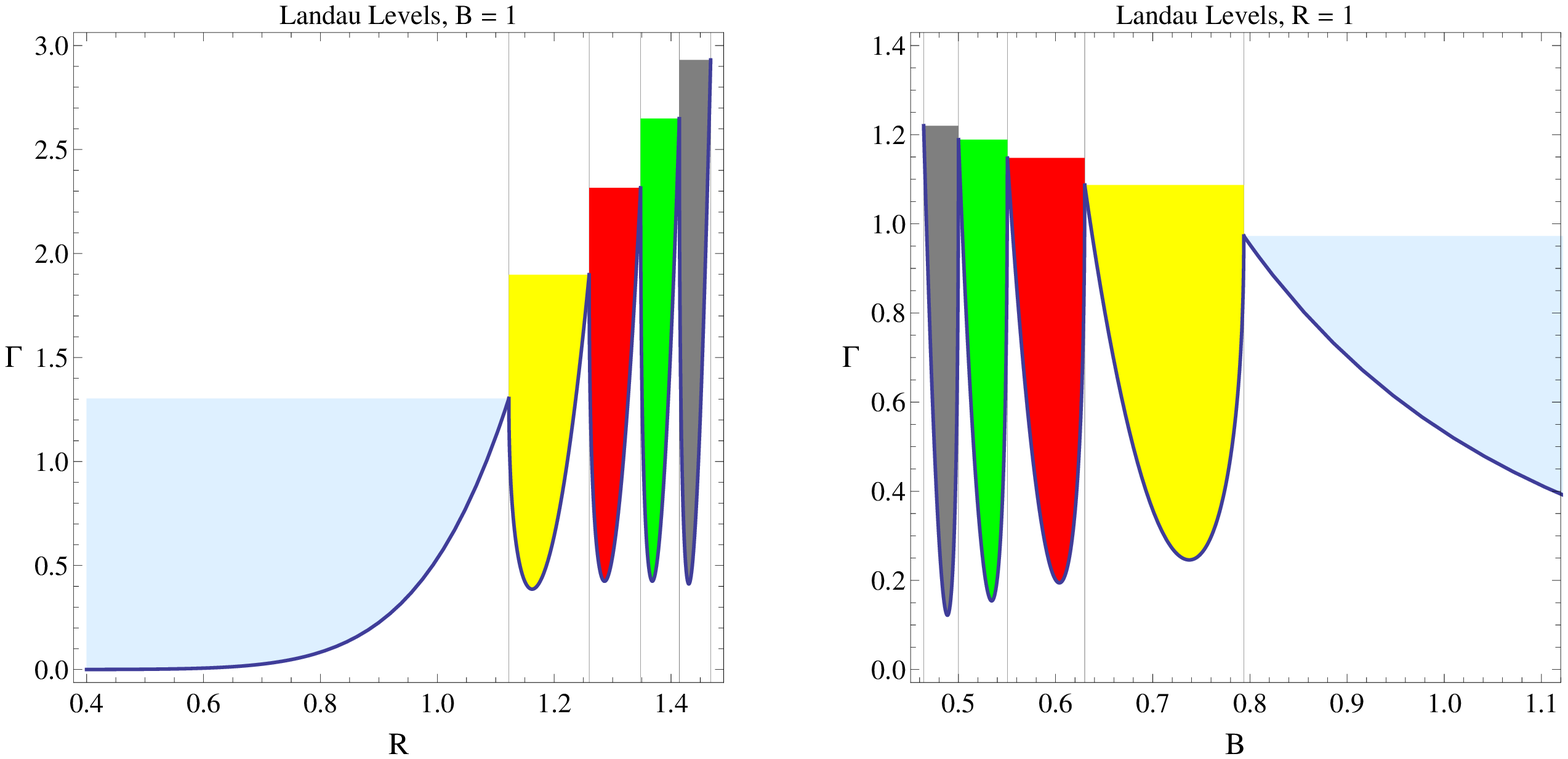}\caption{}
\end{figure}

\newpage

\begin{figure}[ptb]\label{Figure2}
\includegraphics[scale=.6]{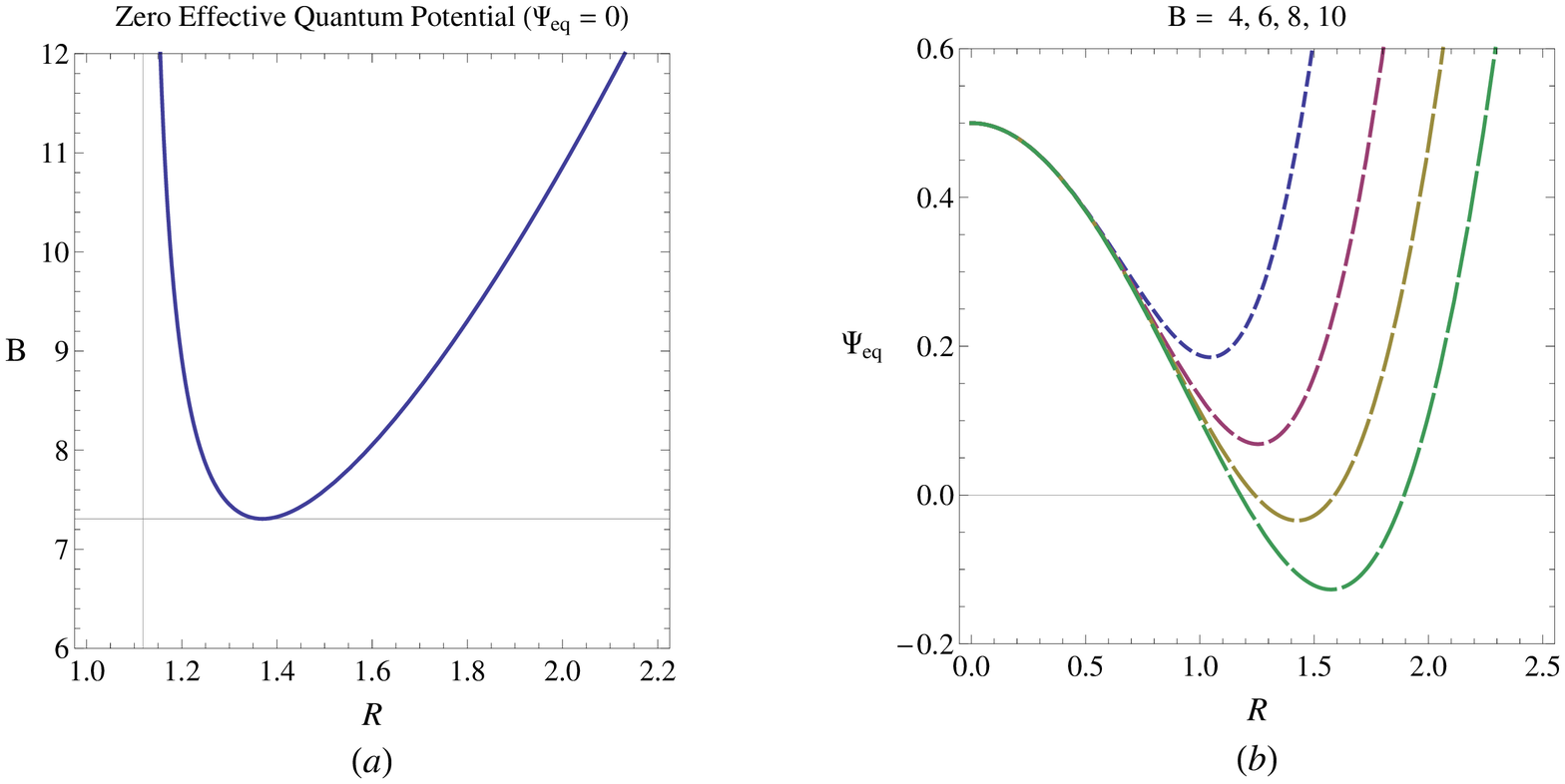}\caption{}
\end{figure}

\newpage

\begin{figure}[ptb]\label{Figure3}
\includegraphics[scale=.6]{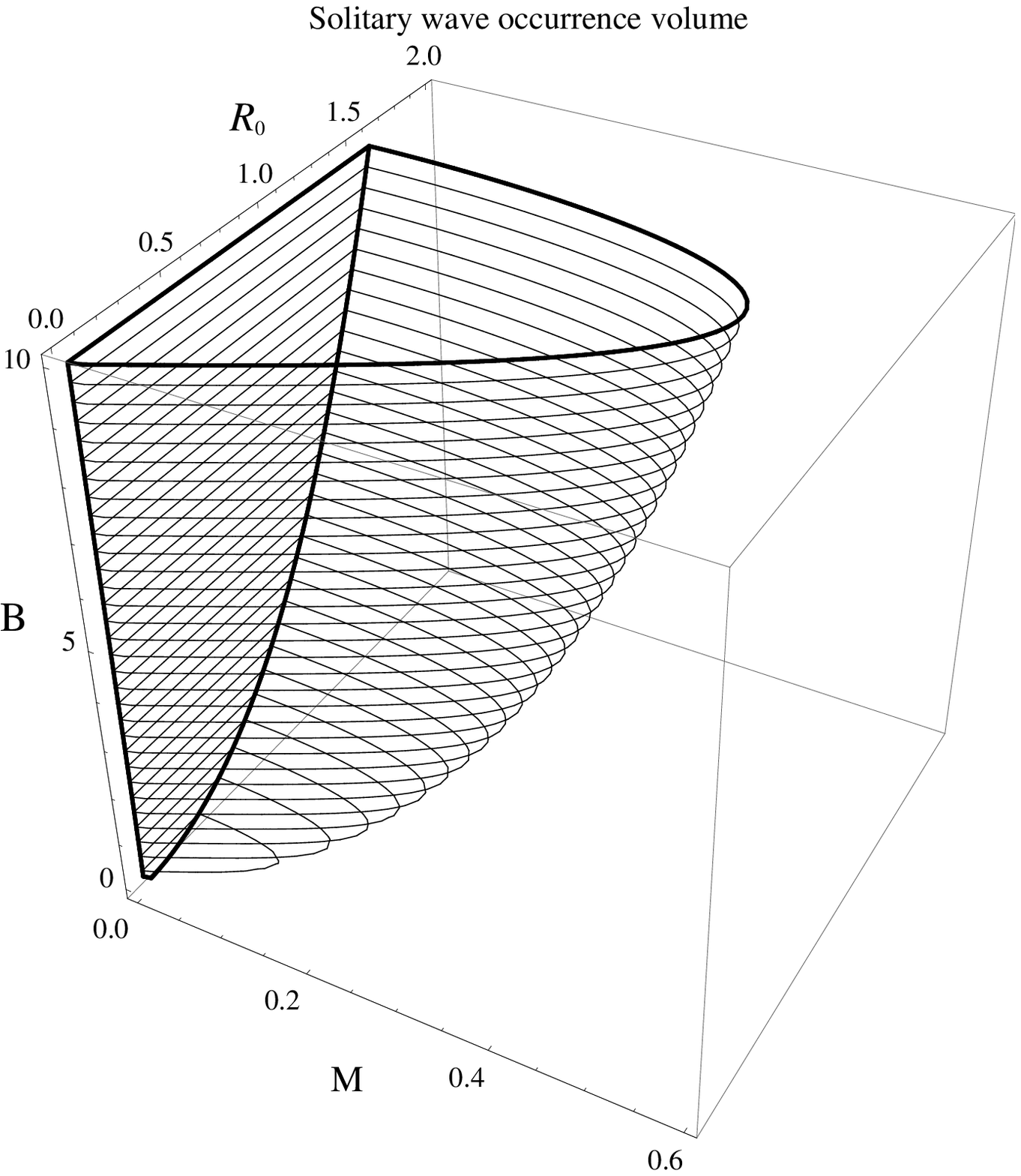}\caption{}
\end{figure}

\newpage
\begin{figure}[ptb]\label{Figure4}
\includegraphics[scale=.6]{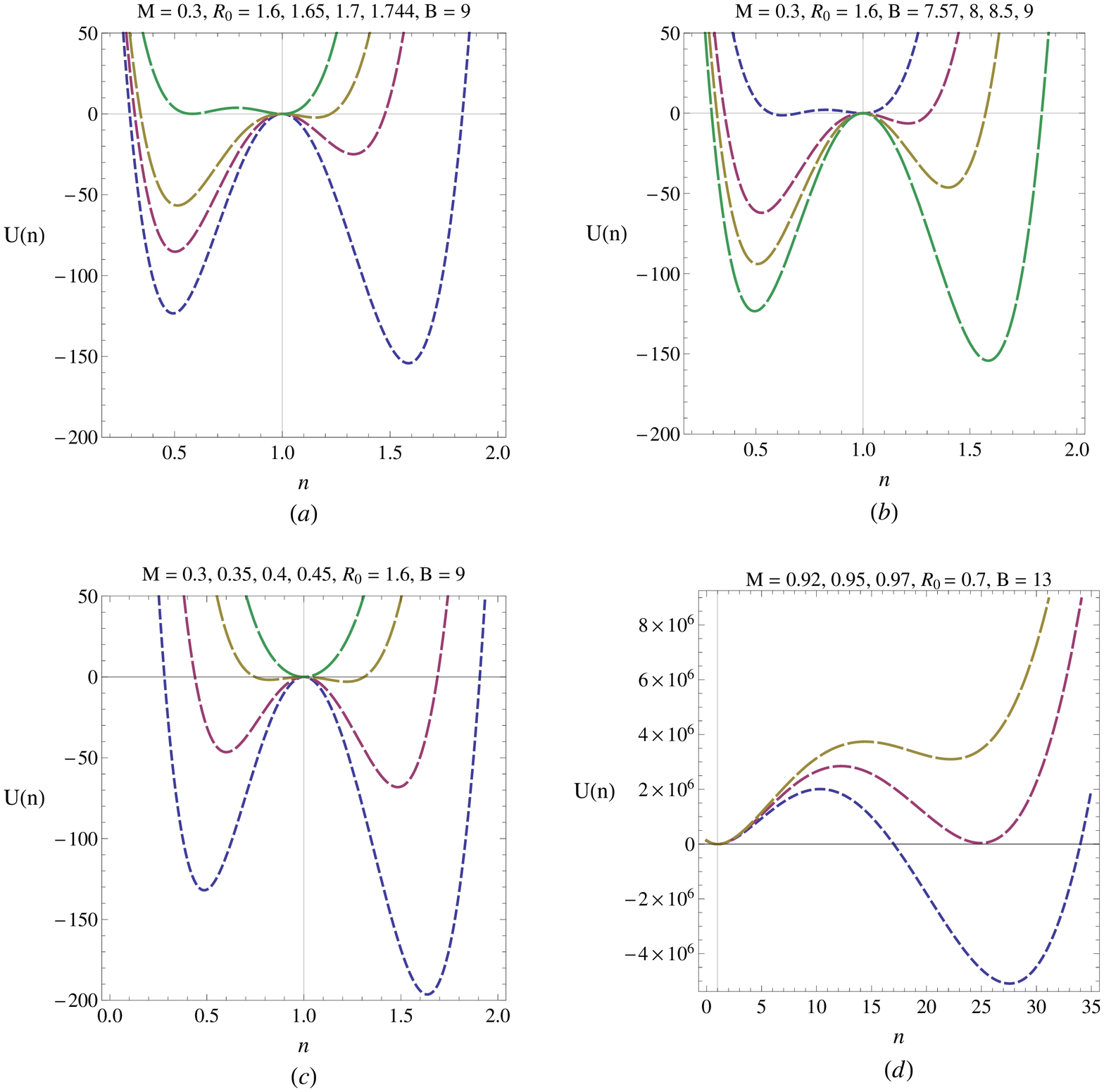}\caption{}
\end{figure}

\newpage

\begin{figure}[ptb]\label{Figure5}
\includegraphics[scale=.6]{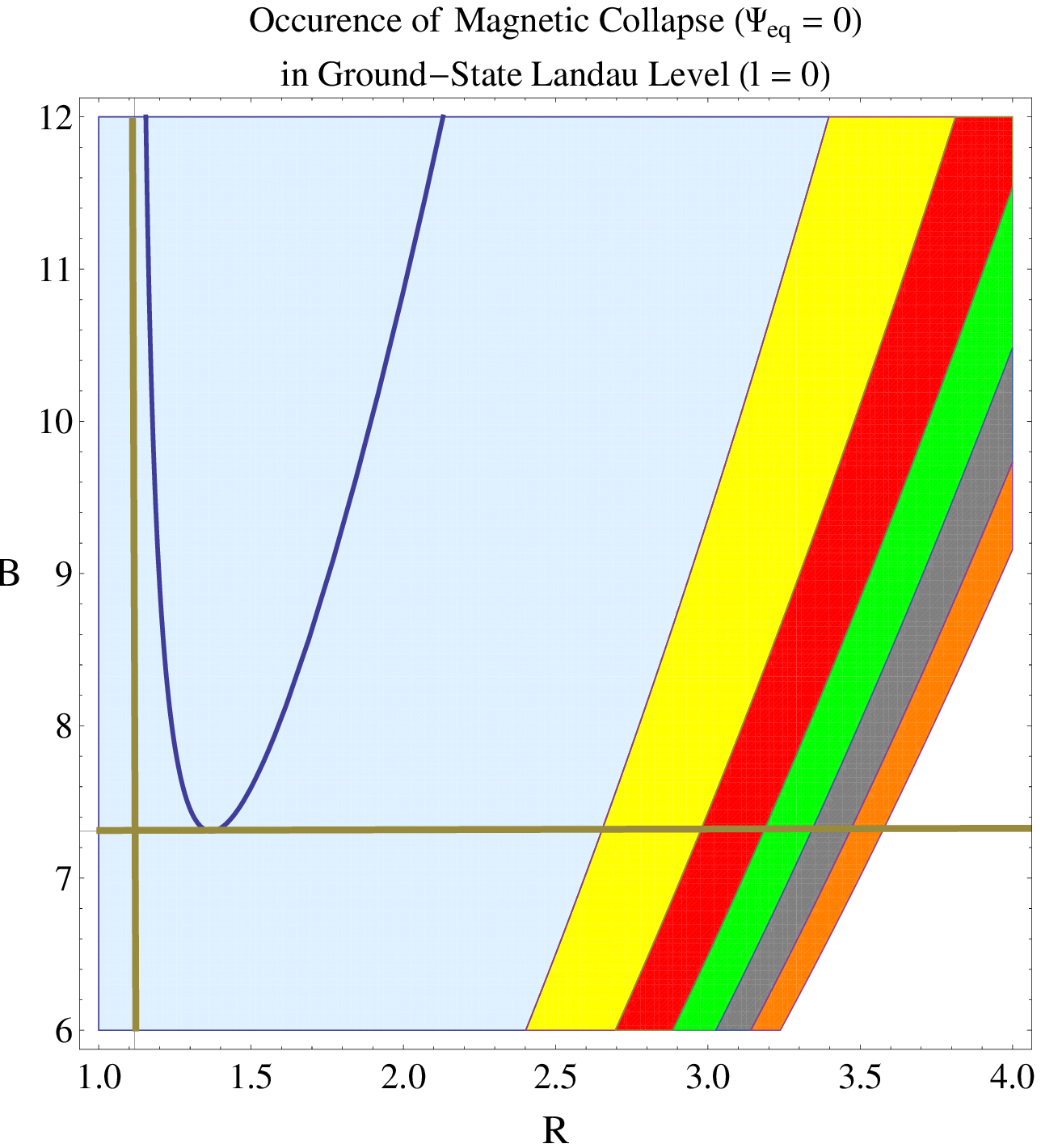}\caption{}
\end{figure}

\end{document}